\documentstyle[array,prl,multicol,aps,psfig]{revtex}
\begin{document}

\title{
A classical model for the negative {\em dc} conductivity of
{\em ac}-driven 2D electrons near the cyclotron resonance}

\author{A. A. Koulakov and M. E. Raikh}
\address{Department of Physics, University of Utah, Salt Lake City,
Utah 84112}
\date{February 23, 2003}

\maketitle


\begin{abstract}
A classical model for {\em dc} transport of two dimensional
electrons in a perpendicular magnetic field and under strong 
irradiation is considered. We demonstrate that, near
the cyclotron resonance condition, and for {\em linear}
polarization of the {\em ac} field, a strong change of the
diagonal component, $\sigma_d$, of the {\em dc} conductivity
occurs in the presence of a {\em weak} nonparabolicity of
the electron spectrum. Small change in the electron
effective mass due to irradiation
can lead to negative $\sigma_d$, while
the Hall component of the {\em dc} conductivity remains practically
unchanged. Within the model considered, the sign of $\sigma_d$
depends on the relative orientation of the {\em dc} and {\em ac}
fields, the sign of the
detuning of the {\em ac} frequency from the cyclotron resonance,
and the sign of nonparabolic term in the energy spectrum.
\end{abstract}

\vspace{5mm}


\noindent{1. {\it Introduction.}}
Recently reported observation\cite{mani,zudov}
of a zero-resistance state,
that emerges upon microwave irradiation of
  a high-mobility 2D electron gas in a weak magnetic field, was
immediately followed by a number of theoretical
papers\cite{phillips,durst,andreev,anderson}, in which the origin
of this state was discussed.
The only microscopic calculation to date\cite{durst}
indicates that, for strong enough radiation intensity,
the diagonal component, $\sigma_d$, of the {\em dc}
conductivity tensor changes sign from the dark value $\sigma_d > 0$
to $\sigma_d < 0$ within certain frequency intervals of the {\em ac}
field, away from the cyclotron frequency and its harmonics.
Negative local value of $\sigma_d$ results in the instability of
the homogeneous current distribution.
In Ref. \onlinecite{andreev} the scenario of
how
the instability might develop into the zero-resistance state
was proposed.

In this situation it seems important to trace the emergence
of negative $\sigma_d$ in an {\em ac}-driven system from the simplest
possible model. Such a model is considered in the present paper.
Obviously, $\sigma_d$ is sensitive to the illumination
only if the Kohn theorem is violated. It is commonly assumed
that the reason for this violation is a random impurity potential.
Here we consider a model, in which the Kohn theorem is
violated due to an intrinsic reason, namely,
due to a weak nonparabolicity of the electron spectrum.
More specifically, we
adopt the following form of the dispersion relation for the conduction
band electrons
\begin{equation}
\label{epsilon}
\varepsilon(p)= \frac{p^2}{2m}\left[1 - \frac{p^2}{2mE_0}\right],
\label{dispersion}
\end{equation}
where $m$ is the effective mass, and $E_0$ is the energy of the order
of the bandgap. The corresponding expression for the velocity reads
\begin{equation}
\label{velocity}
\bbox{v}=\frac{\bbox{p}}{m}\left[ 1- \frac{p^2}{mE_0}\right].
\end{equation}

\noindent{2. {\it dc conductivity.}}
As we will see below, negative $\sigma_d$ emerges when the {\em ac}
field, $\bbox{{\cal E}}\cos \omega t$, is linearly polarized. We choose
the direction of polarization along the $x$-axis. We will also see
that the magnitude and the sign of $\sigma_d$ depend on the
direction of the driving {\em dc} field, ${\bf E}$ (Fig. 1). Denote with
$\theta$ the direction of ${\bf E}$ with respect to the $x$--axis.
Then the  classical equation of motion of an electron in a
perpendicular magnetic field $\bbox{B}$ and driven by
   {\em ac} and  {\em dc} fields is
\begin{equation}
\frac{d {\bbox{p}}}{d t}+\frac{\bbox{p}}{\tau} - \frac{e}{c}
\left[\bbox{v} \times \bbox{B} \right] = e\bbox{E} + e\bbox{\cal E}
\cos \omega t,
\end{equation}
where $\tau$ is the relaxation time.
It is convenient to rewrite this equation introducing a complex
variable ${\cal  P} = p_x + ip_y$. Then it takes the form
\begin{equation}
\label{motion}
\frac{d {\cal P}}{d t} + \frac {{\cal P}}{\tau} - i \omega_c {\cal P} +
\frac{i \omega_c}{m E_0} {\cal P} |{\cal P}|^2 = e E e^{i\theta} +
\frac{e{\cal E}}{2} \left( e^{i\omega t} + e^{-i\omega t}\right).
\end{equation}
Here $\omega_c = eB/mc$ is the cyclotron frequency. We search for a
solution of Eq. (\ref{motion})
in the form
\begin{equation}
\label{complex}
{\cal P}(t)={\cal P}_0+ {\cal P}_{+}\exp(i\omega t) +
{\cal P}_{-}\exp(-i\omega t),
\end{equation}
where ${\cal P}_0 \ll  {\cal P}_{+}, {\cal P}_{-}$ is a small
{\em dc} component proportional to $E$. Near the cyclotron resonance
condition, $\omega \approx \omega_c$, we have
$\vert {\cal P}_{-} \vert \ll \vert {\cal P}_{+} \vert$.
Still we keep the nonresonant term, ${\cal P}_{-}$, to the lowest
order, since the {\em ac} field affects
$\sigma_d$ through this term.
In other words,
the effect of irradiation on $\sigma_d$ emerges
beyond the rotating-wave approximation,
adopted in Ref. \onlinecite{durst}.
Substituting Eq. (\ref{complex}) into Eq. (\ref{motion}), we
obtain the following system of equations for ${\cal P}_{+}$,
${\cal P}_{-}$, and ${\cal P}_0$.
\begin{equation}
\label{pplus}
\left[i(\omega - \omega_c)+\frac{1}{\tau} \right] {\cal P}_{+} +
\frac{i\omega_c}{m E_0}{\cal P}_+|{\cal P}_+|^2 = \frac{e{\cal E}}{2},
\end{equation}
\begin{equation}
\label{pminus}
-i\left( \omega + \omega_c\right) {\cal P}_{-} = \frac{e{\cal E}}{2},
\end{equation}
\begin{equation}
\label{pnul}
\left[ \frac{1}{\tau}-i\omega_c\left( 1-\frac{2|{\cal P}_{+}|^2}{m E_0}
\right)  \right] {\cal P}_{0} + \left[\frac{2i\omega_c} {m E_0}
{\cal P}_{+} {\cal P}_{-} \right]{\cal P}_0^* = eE e^{i\theta}.
\end{equation}
The {\em ac}-induced term $\propto |{\cal P}_{+}|^2$ in the equation
for ${\cal P}_0$ describes the change of the effective mass
caused by irradiation, and has no effect on the {\em dc} transport. The effect
comes from the term $\propto {\cal P}_{+}{\cal P}_{-}$, which
describes the ``rectification'' of the Larmour  motion
due to nonparabolicity.

\begin{figure}
\centerline{
\psfig{file=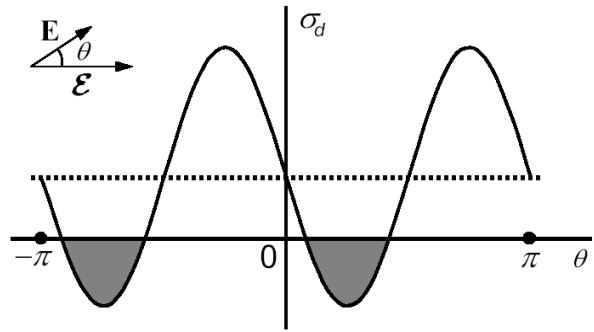,width=3.2in}
}
\caption{The diagonal conductivity as a function of the angle
\protect{$\theta$} between {\em ac} and {\em dc} electric fields,
illustrated in the inset.
The dashed line shows the dark conductivity. The solid curve is a plot of the diagonal conductivity 
under irradiation given by Eq.~(\ref{longitudinal}).
It assumes negative values within the angular intervals shown with gray.}
\label{fig1}
\end{figure}

  From Eq. (\ref{pminus}) we obtain ${\cal P}_{-}
= ie{\cal E}/4\omega_c$. The solution of Eq. (\ref{pplus})
can be formally presented in the form
\begin{equation}
\label{pplussol}
{\cal P}_{+} = \frac{e{\cal E}\tau}{2(1+i\Omega \tau)},
\end{equation}
where 
\begin{equation}
\Omega = \omega - \omega_c + \frac{\omega_c|{\cal P}_{+}|^2}{mE_0}
\label{omega}
\end{equation}  
is the deviation of the microwave  frequency from the
{\em ac}--shifted  cyclotron frequency.
Substituting ${\cal P}_{+}$, ${\cal P}_{-}$ into
Eq. (\ref{pnul}), we find the longitudinal, $p_{\parallel}$,
and transverse,
$p_{\perp}$,  with respect to the applied {\em dc} field, components
of the
drift momentum
\begin{equation}
\label{parallel}
p_{\parallel} = Re \left\{ {\cal P}_0 e^{-i\theta} \right\} = \frac
{eE}{\omega_c^2\tau}\left\{1 - \left[\frac{(e{\cal
E}\tau)^2}{8mE_0}\right]\frac{\Omega \tau \sin 2\theta - \cos
2\theta}{1+(\Omega \tau)^2}  \right\},
\end{equation}
\begin{equation}
\label{perpendicular}
p_{\perp} = Im \left\{ {\cal P}_0 e^{-i\theta} \right\} = \frac
{eE}{\omega_c}\left\{ 1 - \left[\frac{(e{\cal
E})^2\tau}{8mE_0\omega_c}\right]\frac{\Omega
\tau \cos 2\theta + \sin 2\theta}{1+(\Omega \tau)^2}  \right\}.
\end{equation}
The expressions for the diagonal, $\sigma_d$, and
transverse, $\sigma_t$, conductivities immediately follow
from  Eqs. (\ref{parallel}), (\ref{perpendicular}). It is convenient
to rewrite these expressions using the dimensionless parameter
$\delta m/m$, which is the relative correction to the effective mass
due to irradiation
\begin{equation}
\label{dimensionless}
\frac{\delta m}{m} = \frac{|{\cal P}_{+}|^2}{mE_0} = \frac{(e{\cal
E}\tau )^2}{4mE_0\left[1+(\Omega \tau)^2\right]}.
\end{equation}
The above consideration is valid when $\delta m/m \ll 1$. Then we obtain
\begin{equation}
\label{longitudinal}
\sigma_{d} = \frac {ne^2}{m\omega_c}\left[\frac{1}{\omega_c\tau} -
   \left(\frac{\delta m}{2m}\right) \frac{\Omega \tau \sin 2\theta - \cos
2\theta}{\omega_c\tau}  \right],
\end{equation}
\begin{equation}
\label{transverse}
\sigma_{t} = \frac {ne^2}{m\omega_c}\left[1 - \left(\frac{\delta
m}{2m}\right)
\frac{\Omega \tau \cos 2\theta + \sin 2\theta}{\omega_c\tau}  \right],
\end{equation}
where $n$ is the electron concentration.
We see that the {\em ac}-induced correction to the
transverse conductivity is small,
since we assumed that $\vert \Omega \vert \ll \omega_c$ and
$\delta m/m \ll 1$. The latter condition allows us to use
the expansion of $\varepsilon(p)$ given by Eq.~(\ref{dispersion}),
i.e., to neglect the higher--order, in $\varepsilon/E_0$, terms
in the kinetic energy.
Our prime observation, however, is that
with $\delta m /m \ll 1$, the diagonal conductivity,
given by Eq.~(\ref{longitudinal}), becomes {\em negative}
for large
enough detuning from the cyclotron resonance.  The corresponding
condition for negative $\sigma_d$ can be  presented as
\begin{equation}
\frac{\left| \Omega\right|}{\omega_c} > \frac 2{\omega_c \tau}
\left(\frac{m}{\delta m}\right).
\label{thecondition}
\end{equation}
In the clean limit, $\omega_c \tau \gg 1$, this condition is compatible
with $|\Omega| \ll \omega_c$ and $\delta m/m \ll 1$.
The
ratio $\delta m/m$ increases with the intensity
of the {\em ac} field.
Therefore, in order to
meet the condition (\ref{thecondition}),
microwave irradiation should exceed a critical value. For the
intensities above this critical value the diagonal conductivity
assumes negative values
within certain intervals of relative orientation, $\theta$, as illustrated in
Fig.~\ref{fig1}.

It is instructive to analyze the above expressions for diagonal and
transverse conductivities in the limit $\tau \rightarrow \infty$,
where they can be simplified to
\begin{equation}
\label{infty1}
\sigma_{d} = - \left(\frac {ne^2}{m\omega_c}\right)
\left[\frac{(e{\cal E})^2}{8mE_0\Omega\omega_c}\right]
\sin 2\theta ,
\end{equation}
\begin{equation}
\label{infty2}
\sigma_{t} = \frac
{ne^2}{m\omega_c}\left[ 1 -
\frac{(e{\cal E})^2}{8mE_0\Omega\omega_c}
\cos 2\theta \right].
\end{equation}
Remarkably, the relaxation time,
$\tau$, drops out not only from $\sigma_t$, but also
from the diagonal conductivity.  This means that the
the momentum relaxation, necessary for dissipative
transport, is provided by scattering from the microwave field,
coupled to the
translational motion via the nonparabolic term in the dispersion
relation.

\noindent{3. {\it Bistability.}}
In general, $\sigma_d$ in Eq. (\ref{infty1}) is not simply proportional
to the intensity of the {\em ac} field. This is because the effective
detuning, $\Omega$,  also depends on ${\cal E}$, as follows from
Eq. (\ref{omega}). Below we analyze the $\Omega\left({\cal E}\right)$
dependence. For this purpose, we rewrite Eq. (\ref{pplussol}) in the
form
\begin{equation}
z \equiv \frac{|{\cal P_+}|^2}{mE_0} = \frac{|e{\cal
E}|^2}{4mE_0\omega_c^2\left(\delta+z\right)^2}, \ \ \ \delta \equiv
\frac{\omega-\omega_c}{\omega_c}.
\end{equation}
Upon introducing new dimensionless variables 
\begin{equation}
\eta = \frac{|e{\cal E}|^2}{4mE_0\omega_c^2}, \ \ \ s = -z/\delta,
\end{equation}
this equation takes the form 
\begin{equation}
s\left(1-s\right)^2 = -\frac{\eta}{\delta^3}, \ \ \ 
\label{tequation}
\end{equation}
Diagonal conductivity can be expressed through a solution of this
equation as follows
\begin{equation}
\sigma_d = -\frac 12 \sigma_t\delta^2 s \left( s-1\right) \sin 2\theta,
\label{sigmad2}
\end{equation}
where $\sigma_t=ne^2/m\omega_c$ is the transverse conductivity.
Remarkably, Eq.~(\ref{tequation}) yields three solutions for $\delta<0$ and
$\eta/|\delta|^3<4/27$.
Two of these solutions are stable. This fact was first pointed out
by Kaplan \onlinecite{kaplan}, who predicted a hysteresis in the
cyclotron resonance of a free electron,
caused by the relativistic correction to its velocity.
The role of $\tau$ in Ref.~\onlinecite{kaplan} is played by the
radiative friction. This prediction was confirmed experimentally
in Ref.\onlinecite{gabrielse}.

The aspect of nonlinear cyclotron resonance\cite{kaplan},
which is interesting to us,
is how the bistability manifests
itself in the diagonal conductivity, when polarization of the
{\em ac} field is linear.
Two stable solutions of Eq.~(\ref{tequation}) can be obtained
analytically in the limit $\eta \ll |\delta|^3$
\begin{equation}
s_1 = \frac{\eta}{|\delta|^3}, \ \ \ s_2 = 1 + \left( \frac
{\eta}{|\delta|^3} \right)^{1/2}.
\end{equation}
It is seen from Eq.~(\ref{sigmad2}) that they result in two values of
diagonal conductivity
\begin{equation}
\sigma_{d1} = \sigma_t \frac{\eta}{2|\delta|}\sin 2\theta, \ \ \
\sigma_{d2} = -\frac 12 \sigma_t \sqrt{\eta |\delta|} \sin 2 \theta.
\end{equation}
This result is obtained in the limit of infinite $\tau$.
Analysis with finite $\tau$ indicates that the phenomenon of negative
diagonal conductivity vanishes for the second solution,
while $\sigma_{d1}$ remains unaffected.
On the other hand, kinetic energies of these states are
\begin{equation}
\varepsilon_1 = \frac{\eta}{2\delta^2}E_0, \ \ \ \varepsilon_2 = \frac 12 |\delta|E_0.
\end{equation}
Note, that the condition $\eta \ll |\delta|^3$ insures
that $\varepsilon_1 \ll \varepsilon_2$. Our findings can be thus
summarized as follows:
negative $\sigma_d$ corresponds to the lower--energy state,
while in the
higher--energy state the diagonal conductivity remains positive.
Hence, within the model considered in the present paper,
the phenomena resulting from negative diagonal
conductivity should vanish at high enough temperature
due to activation to the state with positive $\sigma_d$.

\noindent{4. {\it Concluding remarks.}} 
We would like to emphasize that the effect of irradiation on {\em dc} transport emerges within our model only for linear polarization of the 
{\em ac} field. As a result, the {\em ac}--induced contribution depends on the relative orientation of the {\em dc} field and 
{\em ac}--polarization.
Simplicity of our model rules out comparison to the experiments \cite{mani,zudov}.
However, we hope that with disorder and interactions present, the real part of electron self-energy
might mimic nonparabolicity of dispersion relation and contribute to the "rectification" of the {\em ac}--driven Larmour motion, 
thus affecting the respose to the {\em dc} field.

The other outcome of our consideration is that in the regime of bistability~\cite{kaplan} 
two stable states of the Larmour rotation exhibit drastically different response to the {\em dc} drive.
When these states have diagonal conductivities of opposite sign, their energies being comparable
may impede the global redistribution of current.



\begin{references}

\bibitem{mani}R. Mani, J. H. Smet, K. von Klitzing, V. Narayanamurti,
W. B. Johnson, and V. Umansky, Nature, {\bf 420}, 646 (2002).

\bibitem{zudov}M. A. Zudov, R. R. Du, L. N. Pfeiffer, and
K. W. West, Phys. Rev. Lett. {\bf 90}, 046807 (2003).

\bibitem{phillips}J. C. Phillips, ArXiv cond-mat/0301254.

\bibitem{durst} A. Durst, S. Sachdev, N. Read, and S. M. Girvin,
ArXiv cond-mat/0301569.

\bibitem{andreev}A. V. Andreev, I. L. Aleiner, and A. J. Millis,
ArXiv cond-mat/0302063.

\bibitem{anderson}P. W. Anderson and W. F. Brinkman,
ArXiv cond-mat/0302129.

\bibitem{kaplan}A. E. Kaplan, Phys. Rev. Lett. {\bf 48}, 138 (1982).

\bibitem{gabrielse} G. Gabrielse, H. Dehmelt, and W. Kells,
Phys. Rev. Lett. {\bf 54}, 537 (1985).



\end{references}
\end{document}